\documentclass{emulateapj}
\usepackage{epsf}
\usepackage{natbib}
\usepackage{apjfonts}
\usepackage{epsfig}

\def\mnras{MNRAS}  % Monthly Notices of the RAS
\def\apj{ApJ}      % The Astrophysical Journal
\def\apjl{ApJL}    % The Astrophysical Journal Letters
\def\apjs{ApJS}    % The Astrophysical Journal Supplement
\def\aap{A\&A}     % Astronomy and Astrophysics
\def\araa{ARA\&A}  % Annual Reviews of Astronomy and Astrophysics
\def\nat{Nature}   % Nature
\def\sci{Science}  % Science
\newcommand{\alphaFet}{[$\alpha$/Fe]}
\newcommand{\alphaFe}{[$\alpha$/Fe]\,}

\newcommand{\FeH}{[Fe/H]\,}

\newcommand{\agefe}{age $-$ [Fe/H]\,}

\newcommand{\magarc}{mag/arcsec$^{2}$}

\shorttitle{Stellar Content of M31's Halo}
\shortauthors{Font et~al.}

\begin{document}

\title{The Stellar Content of Galaxy Halos: A Comparison between  $\Lambda$CDM 
Models and Observations of M31}

\author{
Andreea~S.~Font\altaffilmark{1,7},
Kathryn~V.~Johnston\altaffilmark{1,8},
Annette~M.~N.~Ferguson\altaffilmark{2},
James~S.~Bullock\altaffilmark{3},
Brant~E.~Robertson\altaffilmark{4,9,10},
Jason~Tumlinson\altaffilmark{5},
Puragra Guhathakurta\altaffilmark{6}
}

\altaffiltext{1}{Van Vleck Observatory, Wesleyan University, Middletown, 
CT 06459}
\altaffiltext{2}{Institute for Astronomy, University of Edinburgh, Blackford Hill, Edinburgh, EH9 3HJ, UK}
\altaffiltext{3}{Center for Cosmology, Department of Physics \& Astronomy, University of California, Irvine, CA 92687}
\altaffiltext{4}{Kavli Institute for Cosmological Physics, and Department of Astronomy and Astrophysics, University of Chicago, 933 East 56th Street, Chicago, IL 60637}
\altaffiltext{5}{Yale Center for Astronomy \& Astrophysics, Departments of Physics \& Astronomy, Yale University, P.O. Box 208121, New Haven, CT 06520-8121}
\altaffiltext{6}{UCO/Lick Observatory, Department of Astronomy \& Astrophysics,University of California, Santa Cruz, CA 95064}
\altaffiltext{7}{Institute for Computational Cosmology, University of Durham, Science Laboratories, South Road, Durham DH1 3LE, UK}
\altaffiltext{8}{Department of Astronomy, Columbia University, Pupin Physics Laboratory, 550 West 120th Street, New York, NY 10027}
\altaffiltext{9}{Enrico Fermi Institute, 5640 South Ellis Avenue, Chicago, IL 60637}
\altaffiltext{10}{Spitzer Fellow}

\begin{abstract}

Recent observations have revealed that high surface-brightness,
metal-rich debris is present over large regions of the Andromeda (M31) 
stellar halo. We use a set of numerical models to determine whether 
extended metal-rich debris is expected to exist in galaxy halos formed
in a hierarchical $\Lambda$CDM universe. We identify tidal debris in
the simulations according to the current best surface brightness
detection limits in M31 and demonstrate that bright features in the
outer halo regions usually originate from single satellites, with very
little contamination from other sources due to the low probability of
tidal streams from two overlapping accretion events. In contrast, 
high-surface brightness features in the inner halo often originate
from multiple progenitors. We also compare the age and metallicity 
distribution of the debris with the well-mixed stellar halos in which 
they reside. We find that high-surface brightness tidal 
debris is produced almost exclusively by relatively high mass progenitors 
($M_{*} \sim 10^{7-9} M_{\odot}$) and thus is expected to be of moderate- to 
high-metallicity. Similarly, in our models the smooth inner halo is 
expected to be metal-rich as this region has been built up mainly from 
massive satellites. Our results imply that the stellar populations of 
substructure observed around external galaxies with current techniques 
should not resemble old and metal-poor dwarf spheroidal satellites, nor
the underlying component of the stellar halo. 

\end{abstract}


\section{Introduction}
\label{sec:introduction}

One of the main goals of stellar population studies is to understand the origin of the baryonic components of galaxies. Stellar populations contain a wealth of information $-$ kinematic, age and chemical abundance $-$ that enables the reconstruction of star formation histories of the past and present-day galaxies. More recently, the growing interest in the topic of "near-field cosmology"
  \citep{freeman02} has raised the question of to what extent stellar
  populations can be used to constrain cosmological models.  
In this paper we review recent stellar population data in M31 that appear to challenge hierarchical models of structure formation and compare these data with results of numerical simulations.

Galaxy halos are among the most interesting testing grounds for structure formation theories as they contain imprints from various star formation episodes, starting with those that took place in the earliest formed structures. Patterns in the stellar populations in different regions of a galaxy halo, or between different halos, or even between halos and their neighboring satellite galaxies, can be analyzed and compared with predictions of cosmological models.

The old and metal-poor halo of our Galaxy has long been the prototype for models of galaxy formation \citep{searle78,laird88,ryan91}. Thanks to advances in instrumentational techniques over the last couple of decades it has now become possible to resolve stellar populations in halos of other galaxies within our Local Group: Andromeda (M31) $-$ a galaxy similar in mass and luminosity to our Galaxy \citep{mould86,reitzel98,durrell01,durrell04,ferguson02,ferguson05,brown03,brown06a,brown06b,brown07,ibata07}, and various satellite galaxies, such as the Magellanic Clouds, M33 and most of the dwarf spheroidals \citep{harris01,mcconacchie06,mighell96,hurley-keller98,gallart99,harbeck01,dolphin02,martin06}. Whereas the M31 and Milky Way halos exhibit similar stellar compositions (an underlying smooth, metal-poor component and a few metal-rich tidal structures), some local dwarf galaxies are quite different from the large halos. Some of these dwarfs have stellar populations that are on average younger and more metal-rich than the halo of our Galaxy, but overall display a great diversity in their stellar populations \citep{mateo98,tolstoy04,dolphin05}. 

Recently, it has also become possible to analyze the stellar populations in some of the faintest galactic substructure, namely in the tidal debris. In contrast with the variety of populations displayed by neighboring dwarf galaxies, the streams discovered so far are striking in their uniformity. Unlike the predominantly old and metal-poor populations of nearby dwarf spheroidal satellite galaxies, the streams discovered so far are predominantly metal-rich with significant intermediate age components \citep{ferguson05,brown06b}. Moreover, the debris has distinct ages and metallicities even in comparison with the surrounding halo in which it resides. In the Milky Way, the Sagittarius stream $-$ the largest substructure discovered in the halo \citep{ibata97,majewski99,majewski03} $-$ has significantly younger stars than the stellar halo and an average metallicity that is a few dex higher \citep{yanny00,ivezic00,delgado04,bellazzini06,monaco06,belokurov06}. An even larger stream dominates the halo of M31 \citep{ibata01}. This so-called ``giant southern stream'' (GSS) has a mean metallicity of \FeH $=-0.5$, higher by about $0.2$~dex than the inner M31 halo \citep{ferguson02,guhathakurta06}, and by about $1$~dex than the metal-poor M31 halo component \citep{chapman06,kalirai06}. The stream also has a larger fraction of intermediate-age stars \citep{brown06a}. Other pieces of debris discovered in M31 include the ``NGC205 loop'', the ``G1 clump'', the ``North-East shelf'', the ``Northern spur'' \citep{ferguson02,ferguson05,mcconacchie04}, etc.  $-$ the latest inventory done by \citet{ibata07} in M31 lists about a dozen or so tidal features. Detailed studies of different features in the color-magnitude diagrams in these regions suggest that the metallicities are uniform and high, \FeH $\sim -0.4$ \citep{ferguson05,faria07}. 

Are all the data consistent with the hierarchical model for structure
formation? At first sight, the diversity of stellar populations in the
halos, tidal streams and surviving satellites is suggestive of a
stochastic assembly. However, some of the data have been given another
interpretation. For example, it has been argued that the {\it differences} in chemical abundances between neighboring satellites and the Milky Way halo exclude these satellites from the list of possible halo progenitor types \citep{venn04}. At the same time, one can use the {\it uniformity} of chemical abundances in M31's spheroid as an argument against the hierarchical scenario: if the halo formed from different progenitors, why is the debris so uniform in metallicity \citep{ferguson05}? Other recent observations claiming to challenge the hierarchical scenario use as arguments the apparent lack of very metal-poor stars in some of the surviving dwarf galaxies but not in the stellar halo \citep{helmi06}, and the metal-poor component of the M31 extending up to large distances in an apparently homogeneous fashion \citep{ibata07}.
 
From the above arguments, it is clear that better theoretical expectations are needed in terms of the distribution of stellar populations in dwarf galaxies, stellar halos and in the debris. In a recent series of papers we have begun to compare the results of theoretical models with available data, using as a basis a set of simulations of stellar halos formed in a $\Lambda$CDM cosmology. We found that the observed differences in \alphaFe abundances between the surviving satellites and the halo arise naturally in the $\Lambda$CDM framework \citep{robertson05,font06a}. We have also shown how the phase-space structure of current debris can be used to recover the merger histories of $\Lambda$CDM stellar halos \citep{font06b,johnston07} or identify halos whose merger histories are atypical for this cosmology \citep{johnston07}. 

In this paper we make a direct comparison between between the stellar
populations in our models and those in the halo of M31, down to a level
of surface brightness similar to current detection limits. We choose
M31 because this galaxy has benefited in recent years from a systematic
study of its stellar populations, using either deep HST photometry,
deep field spectroscopy or dedicated wide-field surveys
\citep{ferguson02,ferguson05,brown03,brown06a,brown06b,brown07,guhathakurta05,guhathakurta06,irwin05,gilbert06,gilbert07,kalirai06,ibata07,tanaka07}.
As a result, some aspects of the stellar populations in the M31 halo
are known better now than those in the Milky Way, where observations
target mainly the Solar neighborhood. In addition, M31 exhibits a wide
spread distribution of metal-rich debris throughout its halo. Our aim
is to investigate whether this distribution is consistent with the
hierarchical merger model. We use the age $-$ metallicity relation
(AMR) as an indicator for the formation of stars in our systems (both
in the smooth stellar halo and in the tidal streams). We then compare the AMRs
from our simulations with those obtained  from the available stellar
population data in M31.

We also seek to constrain the origin of the bright, metal-rich debris
in stellar halos (and in particular in M31). In observations it is
often difficult to disentangle multiple stream components in the
debris. In some cases the stars may belong to overlapping orbits of a
single satellite (as it is now becoming evident in the case of the
Sagittarius stream in the Milky Way). In others, the ``debris'' could
be just the perturbed or warped outer disk $-$ as it may be the case
for some of the recent detections near the edges of M31 and Milky Way's
disks \citep{momany06,faria07}. When available, kinematic data can be
used to separate the streams in phase-space \citep{helmi00}, and the
modelling of orbits can further connect stellar features that are
physically separate on the sky. Additionally, it is conceivable,
although not yet proven, that stellar populations can be used as
``tags'' for the individual satellites from which they originate. In
this paper, we use a statistical sample of bright tidal debris fields
to provide a quantitative prediction for their origin. We estimate the 
average number of satellites per field with significant contributions
to the bright debris, both in the inner and outer simulated halos. We also
analyze the average metallicity of the bright debris and found that it
is in good agreement with the observations in M31. 

The paper is structured as follows: In section \ref{sec:models} we
summarize the models. In section \ref{sec:results} we present the AMRs
of the main galaxy halos (\S \ref{sec:theor_halos}) and of tidal
streams (\S \ref{sec:theor_streams}). In section \ref{sec:highsb} we
investigate the origin of the highest surface brightness features (\S
\ref{sec:origin}),  their typical ages and metallicities (\S
\ref{sec:agefe_fields}). In Section \ref{sec:comp_obs} we make detailed 
comparisons with available M31 data, for the halo (\S
\ref{sec:comp_obs_halo}) and the GSS (\S \ref{sec:comp_obs_stream} and \S \ref{sec:sfr_orbit}). In Section \ref{sec:concl} we conclude.

\section{Models}
\label{sec:models}

The set of stellar galaxy halos used here has been constructed by \citet{bullock05} and it comprises 11 dark matter halos of mass $M_{vir} = 1.4 \times 10^{12} M_{\odot}$ formed in a $\Lambda$CDM Universe ($\Omega_{m}=0.3,  \, \Omega_{\Lambda}=0.7, \,  h=0.7,$ and $\sigma_{8}=0.9$). For our discussion we will also refer to these as $L_{*}$ galaxy halos. 

Full details about the set-up of the simulations can be found in the
previous papers \citep{bullock05,robertson05,font06a}. Here we just
briefly summarize the main characteristics of the models. The merger
history of the galaxies is based on an Extended Press-Schecter
formalism \citep{lacey93} and the evolution of each satellite is
followed with a numerical simulation after accretion onto the main
halo. The baryonic components are modeled with semi-analytical
prescriptions. The infall of cold gas into each satellite is followed
from the epoch of reionization to the time of its accretion onto the
main halo. The gas mass is translated into a stellar mass through a
simple star-formation prescription and the stellar profile is assumed
to follow a \citet{king62} model. Star formation ceases soon after
accretion onto the main halo when the gas is assumed to be lost due to
ram-pressure stripping. The chemical evolution of the accreted
substructure is followed with the method of \citet{robertson05}, which
takes into account the enrichment from both Type II and Type Ia
supernovae and models feedback as metal-enriched winds from both
supernovae and intermediate mass stars. The models evolve with
redshift, but are calibrated only at redshift $z=0$ to match the 
stellar mass $-$ metallicity relation, $M_{*} - Z$, and the stellar
mass $-$ circular velocity relation, $M_{*} - $v$_{\rm circ}$, for
Local Group dwarfs.

When evolved to the present day, the models match the main physical properties of the two main galaxies in the Local Group and of their progenitors: the number of bright ($> 10^5 L_\odot$) surviving satellites per main galaxy, and the masses, density profiles and total luminosities of the stellar halos are similar to those found in the Milky Way and M31. The average metallicities of our simulated $L_{*}$-type galaxy halos range between $-1.3$ and $-0.9$. The model also retrieves the differences in the \alphaFe abundances between stellar halos and the surviving satellites. More details about the spatial distribution of chemical abundances (\FeH and \alphaFet) in these stellar halos can be found in \citet{font06a,font06b}.

\section{Results}
\label{sec:results}

\subsection{Predicted properties of stellar halos}
\label{sec:theor_halos}

The AMR  of a stellar population is a useful tool for gaining insight into the chemical evolution of the system.  In a closed box system, it is expected that the oldest populations will be the most metal poor whereas the youngest ones the most metal rich. The slope of the AMR can vary, being generally steeper for the initial stages of the evolution. The AMR will be monotonic only for isolated systems that formed monolithically \citep{pagel75,binney98}, while systems that formed by accretion of satellites will display a composite functional form, with multiple metallicity components for the same age. In this section we use our models to investigate to what extent the AMR of a stellar halo can be used to place constraints on the galaxy mass assembly history.

\begin{figure}
\centering
\includegraphics[width=9cm]{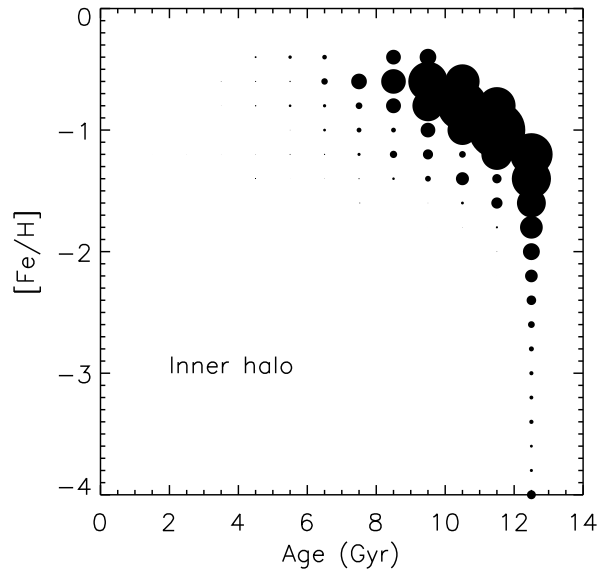}
\caption{\label{fig:agefe_master}{AMR for the average of all 11 halos in our sample. Only the inner $50$~kpc regions of the halos have been considered here. The area of symbols varies with the mass fraction of stellar populations to the total mass in the field. (No population contributes, by mass, more than $25\%$ to the halo.)}}
\end{figure}

\begin{figure}
\centering
\includegraphics[width=9cm]{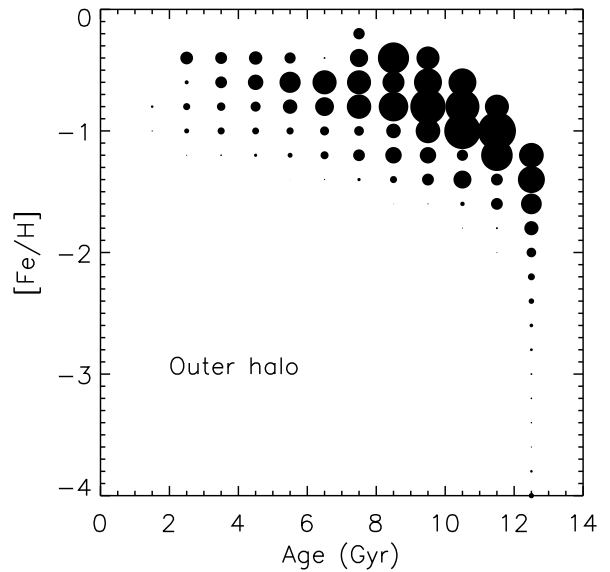}
\caption{\label{fig:agefe_inner_outer}{AMR for the outer region ($50<R$~kpc) of the averaged simulated halo. Symbol areas correspond to the stellar mass fraction contributed by the population, as in Figure \ref{fig:agefe_master}. The fraction of each population is normalized to the stellar mass of the entire outer halo region.}}
\end{figure}

\begin{figure*}[t]
\includegraphics[width=16cm]{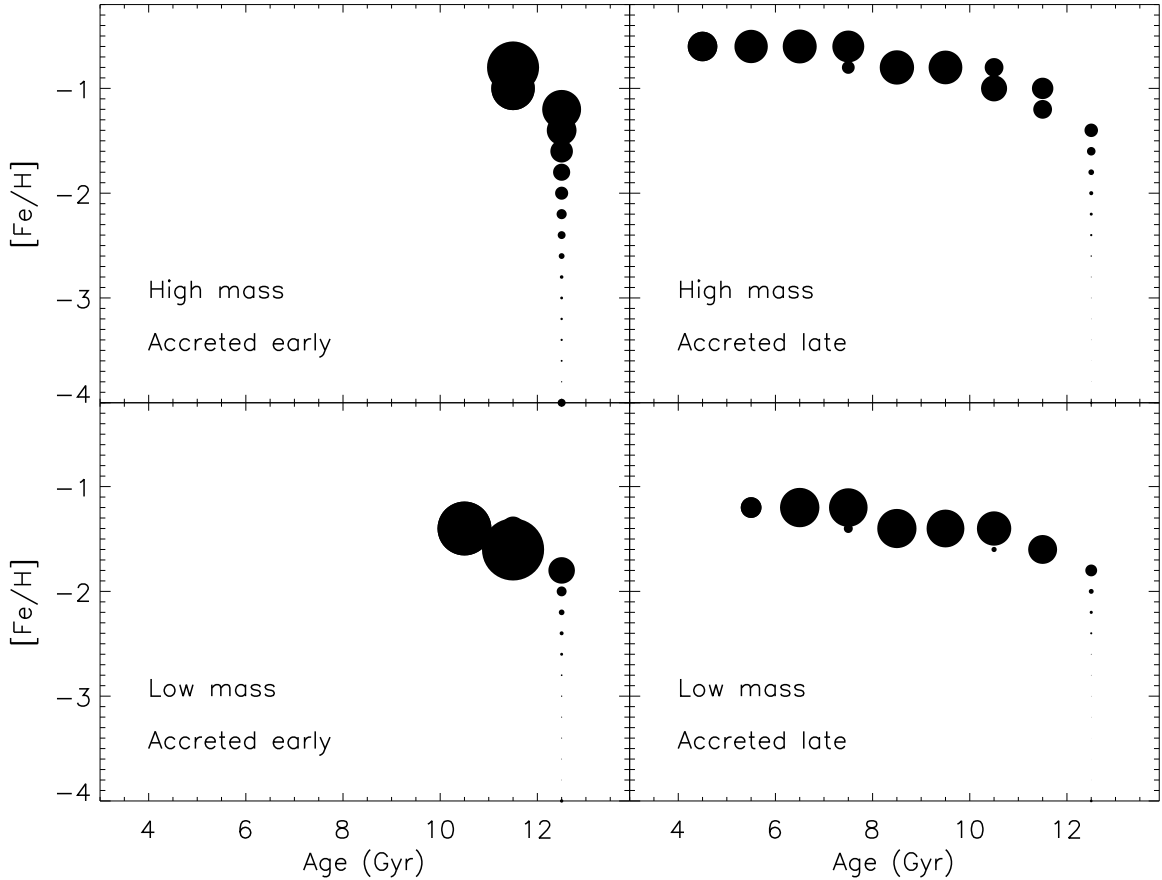}
\caption{\label{fig:agefe_streams}{AMRs of individual satellites of various masses and accreted at different times. The low mass satellites chosen here have $M_{*} \sim 10^{6} M_{\odot}$, while the high mass ones have $M_{*}$ of a few $10^{8} M_{\odot}$. The accretion times are of $1-2$~Gyr for the early events and $6-7$~Gyr for the late ones. Symbol areas correspond to the stellar fraction contributed by the population, as in Figure \ref{fig:agefe_master}. The stellar mass fractions are normalized to the total stellar mass of each satellite.
}}
\end{figure*}

Figure \ref{fig:agefe_master} shows the AMR of our ``typical'' stellar halo, constructed by averaging over all 11 halos in our sample. Each symbol represents a cell in the 2-dimensional \agefe  grid where there are any stars\footnote{We henceforth  consider all stars in a given grid \agefe cell to  be a distinct stellar population.}, with the area of the symbol corresponding to the mass fraction contributed by the stellar population to the total stellar mass of the averaged halo. Only the inner $50$~kpc of the stellar halos are considered for this plot.

The AMR in Figure \ref{fig:agefe_master} has a number of distinctive
characteristics: 1) the AMR has a single dominant sequence rather than
being composed of multiple ones; 2) the sequence is representative of
satellites that are massive, i.e that enrich to \FeH $>-1$ values in
just a few Gyrs; and 3) the ages of the stars are typically old,
$>8$~Gyr. These characteristics are a direct consequence of the merger
histories of inner halos assumed for our $\Lambda$CDM models, and which
have been described in more detail in previous papers
\citep{bullock05,font06a,font06b}. The dominant AMR sequence occurs
because the bulk of the inner halo is formed from a few massive
satellites ($10^{7}-10^{8} M_{\odot}$), even though the main halo
accreted satellites with a broader range of stellar masses ($M_{*} \sim
10^{4}-10^{8} M_{\odot}$). In addition, massive satellites sink more
rapidly into the center due to a more effective drag of dynamical
friction. The major contribution to the total stellar mass of massive
satellites also explains the clear signature of rapid metal enrichment
(more details about the AMRs of individual satellites are given
below). Few stars have ages $< 8$~Gyrs, and this is because early
events dominate the inner halo accretion history.

Spatial variations in the stellar populations properties across hierarchically-formed halo are also of observational interest. Figure \ref{fig:agefe_inner_outer} shows the AMR of the outer regions of halos ($R>50$~kpc),  using the stellar population data of the same averaged halo as in Figure \ref{fig:agefe_master}.

The larger spread in ages in the outer halo is a consequence of the late assembly of this region compared with the inner one \citep{bullock05,font06a}. Given the evolution with time of the typical mass of the accreted systems, the outer halo is formed from satellites with higher masses, $M_{*} \sim 10^{6}-10^{9} M_{\odot}$, than the inner halo. Of note, is the presence of very massive satellites in the outer halo, $M_{*} \sim 10^{9} M_{\odot}$, which extend the outer halo AMR towards \FeH $\sim 0$ values.

Our results indicate that an averaged sample of the outer halo should display a significantly larger fraction of intermediate-age versus old stars than the inner halo. The metallicity distribution function should also have a broader peak in the outer halo. 

\subsection{Predicted properties of individual satellites and streams}
\label{sec:theor_streams}

By construction our simulated satellites have a monotonic chemical enrichment. This limitation results from a number of simplifying assumptions $-$ for example, the merger history of a satellite is approximated with a general formula \citep{wechsler02} and not followed stochastically, and the gas and metals are assumed to be instantaneously mixed after each star formation episode. The predictive power of our models for individual satellites (and streams) is therefore restricted. The individual AMRs will underestimate the scatter in \FeH for stellar populations of the same age. However our models are fairly reliable in matching the mean metallicity of satellites and the overall AMR sequence.

Figure \ref{fig:agefe_streams} shows four cases of individual satellites in the range of stellar masses and accretion times that are typically found in our simulations. Despite the limitations in modelling the entire scatter, some trends are obvious. For example, satellites have a smaller \FeH spread at a fixed age than the outer halo but the spread is comparable to that of the inner halo.  Also, the overall shape and maximum \FeH of the inner halo AMR are similar to those of massive satellites ($M_{*} > 10^{8} M_{\odot}$).

The AMR of the outer halo contains individual AMR sequences that are
more clearly separated (Figure \ref{fig:agefe_inner_outer}) than those
in the inner region of the halo (Figure \ref{fig:agefe_master}). As
each AMR sequence corresponds to a single satellite or a group
of satellites with similar properties, this suggests that using AMRs as
a tool to disentangle the accretion histories of stellar halos may be
more feasible in the outer rather than in the inner regions of the halos.

\subsection{The Extended, High Surface Brightness Debris}
\label{sec:highsb}

\subsubsection{Origin}
\label{sec:origin}

What is the origin of the bright, metal-rich debris extending over large areas in the M31 halo?  Was a single satellite responsible for polluting the M31 halo with metal-rich stars or were they several? Here we use our models to constrain the formation history of this debris, assuming that it formed entirely from merging events.

\begin{figure}[t]
\centering
\includegraphics[width=9cm]{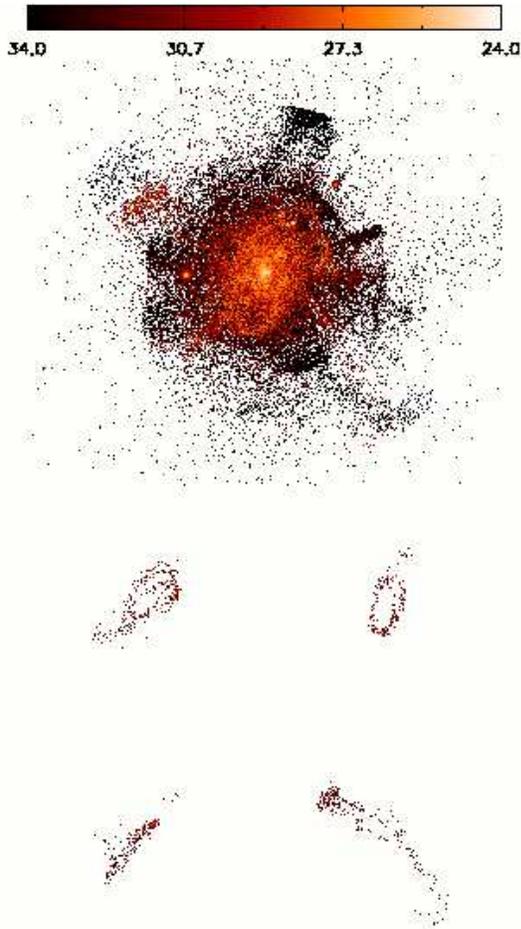}
\caption{\label{fig:short}{
{\it Top:} A stellar halo mapped in surface brightness down to $34$ \magarc (this is halo H2 in \citet{font06b}). The region shown here extends $150$~kpc on the side. {\it Middle and bottom:} Four examples of kinematically cold streams surviving at present time in the halo above. The streams have been shrunk to a quarter of their size, for illustration purposes.
}}
\end{figure}

The upper panel in Figure \ref{fig:short} shows the surface brightness map of a stellar halo with M31-like properties. The map shows only the surface brightness down to the limit of $\mu_{V} =34$ \magarc, which is roughly the deepest limit achieved currently by wide field surveys of the M31 \citep{ferguson02,irwin05,ibata07}. The bottom panels in Figure \ref{fig:short} show the surface brightness maps of the main four kinematically cold streams surviving at the present time in the simulated halo.

Down to the chosen surface brightness cut, the spatial distribution of the simulated debris is similar to what has been found so far in M31 and in the Milky Way: there are a few bright kinematically cold streams (the counterparts of the Sagittarius stream, the GSS or the NGC205 loop), the rest of the debris being in the form of much fainter streams or diffuse overdensities. 

We define as ``high surface brightness'' all debris that is as bright (or brighter) than the current detections of substructure $-$ $\mu_{V} \sim 30$ \magarc \, for the GSS in M31 \citep{ibata01}. We therefore select for our analysis all fields which have $\mu_{V} \le 30$ \magarc \, and that are located on the projected area of a halo along the line of sight. These mock fields have the same area as the pixel size area of the halo map in Figure \ref{fig:short}, i.e. $0.4 \times 0.4$ kpc. The depth of the field along the line of sight is of the same order as the virial radius of the main galaxy, i.e. $\simeq 300$~kpc. The total surface brightness in each pixel is obtained by integrating over all populations in the field.

\begin{figure}
\centering
\includegraphics[width=9cm]{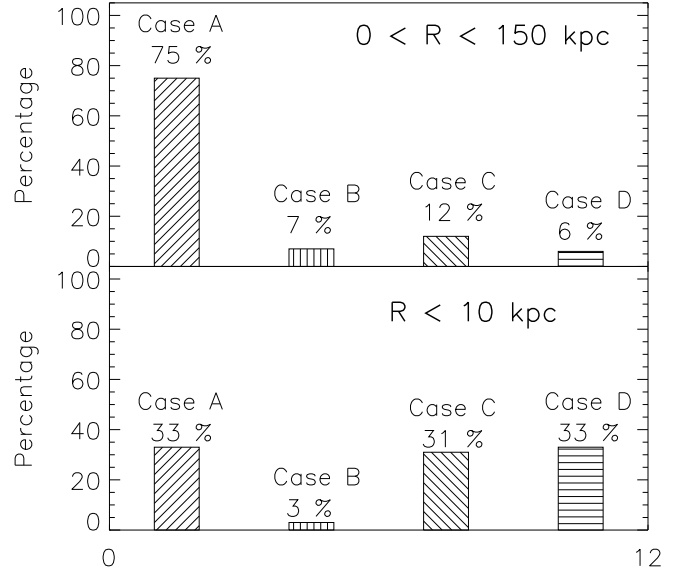}
\caption{\label{fig:hist}{The origin of high surface brightness fields ($\mu_{V} \le 30$ \magarc): a single source (case A), two dominant sources (case B), multiple less dominant sources (case C) and diffuse material originating in many minor sources (case D). {\it Top panel:} All bright fields across the halo are selected; {\it Bottom panel:} Only the bright fields in the inner $R<10$~kpc of the halo are selected.}}
\end{figure}

We then divide all bright fields into four disjoint cases, based on the formation history of the stellar overdensity in that field: 

\vskip 8pt

- case A: The overdensity originates mainly in a single satellite. The condition we impose is that the satellite contributes a mass fraction $f >80\%$ of the total stellar mass in the field;

- case B: The overdensity is formed from contributions of two main satellites. The condition is that these satellites combined contribute more than $80\%$ of the stellar mass in the field;
 
- case C: The overdensity is formed from one or more satellites with
  less significant contribution(s). Specifically, we require that there is
  at least one satellite contributing $30\% < f < 40\%$, but no
  satellite contributes more than $40\%$ to the total stellar mass in
  the field;

- case D: The overdensity is formed from a multitude of minor satellite contributions. No satellite is contributing more than $30\%$ of the total mass in the field.
\vskip 8pt

The top panel of Figure \ref{fig:hist} shows the frequency of each case among the bright fields over the entire area of the stellar halo. Cases in which the overdensity originates in a single satellite (A) occur about $70\%$ of the time, while cases B, C and D occur less frequently, each at about $10\%$ of the time. The actual percentages vary slightly from halo to halo and depend on the imposed surface brightness limit -- for example, the frequency of case A increases if we select an even brighter debris. However, case A consistently dominates across the halos. The distribution of these cases begins to reverse in the inner regions of the halos, where phase-mixing is more important. The bottom panel in Figure \ref{fig:hist} shows the same information as the top panel, but for the inner $10$~kpc of the halo. Here, the dominating cases are those in which the debris originates in three or more satellites. 

These results have a number of implications for the interpretation of high surface brightness overdensities in galaxies. For example, if the overdensity is relatively distant from the center of the halo, $R>10-20$~kpc, it is likely to originate in a single satellite, with little contamination from other satellites in the field. This result holds regardless of whether the feature is kinematically cold or if it is diffused in phase-space. For the inner halo, only a third of the bright features are created by a single satellite, the majority of them originating from superposition of streams from three or more satellites (cases C and D).

\subsubsection{Metallicity}
\label{sec:agefe_fields}

The existence of the metallicity - luminosity relation
\citep{garnett87,skillman89,brodie91,zaritsky94} and the corresponding 
metallicity - mass relation \citep{lequeux79,tremonti04,erb06} for galaxies
indicates a connection between the nature of the satellite progenitors
and their metallicities. We might expect  a corresponding connection to
exist between the surface brightness of the current debris and its
overall metallicity. Given this expectation, it is useful to examine
whether high surface brightness debris in our models (as analogue of
the M31 observations) indeed originates in satellites of a typical
stellar mass, and if so, what is the predicted average metallicity of
the debris.

In our models, satellites relatively massive, $M_{*} \sim 10^{7-8}
M_{\odot}$, and accreted $10-12$~Gyr ago result in sufficiently bright
debris to be detected, $\mu_{V} \sim 30-32$ \magarc \,
\citep{johnston07}. As the surface brightness of the debris depends on
both the mass of the progenitor, and on the time of accretion,
satellites with $M_{*} \sim 10^{7-8} M_{\odot}$ accreted less than
$10$~Gyr ago are much brighter, $\mu_{V} \sim 28-30$
\magarc. Satellites with stellar masses greater than $10^{8} M_{\odot}$
assembly later and are therefore accreted later ($< 8$~Gyr ago), 
resulting in debris with the highest surface brightness, $\mu_{V} \sim
26-28$ \magarc \, \citep{johnston07}. Therefore, the average
metallicities of high surface brightness fields originating from a
single satellite are expected to range between $-1.4$ and $0$ (which
are average values for satellites with stellar masses of $10^{7}$ and
$10^{9} M_{\odot}$, respectively; values are quoted at the time of accretion).

A high surface brightness in a given field can be achieved not
only from the contribution of a single relatively massive satellite,
but also from the superposition of debris originating in several
massive satellites (eg. if $80\%$ of field stars originating in a
$10^{7-8} M_{\odot}$ satellite reach the detection limit, i.e a case A,
then a combination of two $40\%$ fractions originating in two separate
$10^{7-8} M_{\odot}$ satellites are equally bright, i.e a case B). In
the case of superposition of massive satellites, the average
metallicity of the high surface brightness field is similar to the
average metallicity of the individual satellites, i.e. the fields are
expected to be metal-rich.

As we will discuss below, most of the observed debris in M31 {\it is}
metal-rich. The average \FeH in the GSS is $\sim -0.5$
\citep{ferguson02,brown06b} and other debris throughout the M31 halo
have similarly high metallicities, \FeH $\sim -0.4$
\citep{ferguson05,faria07}. Additionally, \citet{gilbert07} show that
the M31 ``spheroid'' fields and several fields along the minor axis are
consistently metal-rich, with and without the cold GSS component. The
lesson learnt from our models is that the metal-rich debris currently
observed in the inner halo of M31 may not have a single common
progenitor, and moreover, each individual inner halo field may contain 
metal-rich components from multiple sources. 

\section{Applications to Observations}
\label{sec:comp_obs}

\begin{figure*}[t]
\includegraphics[width=16cm]{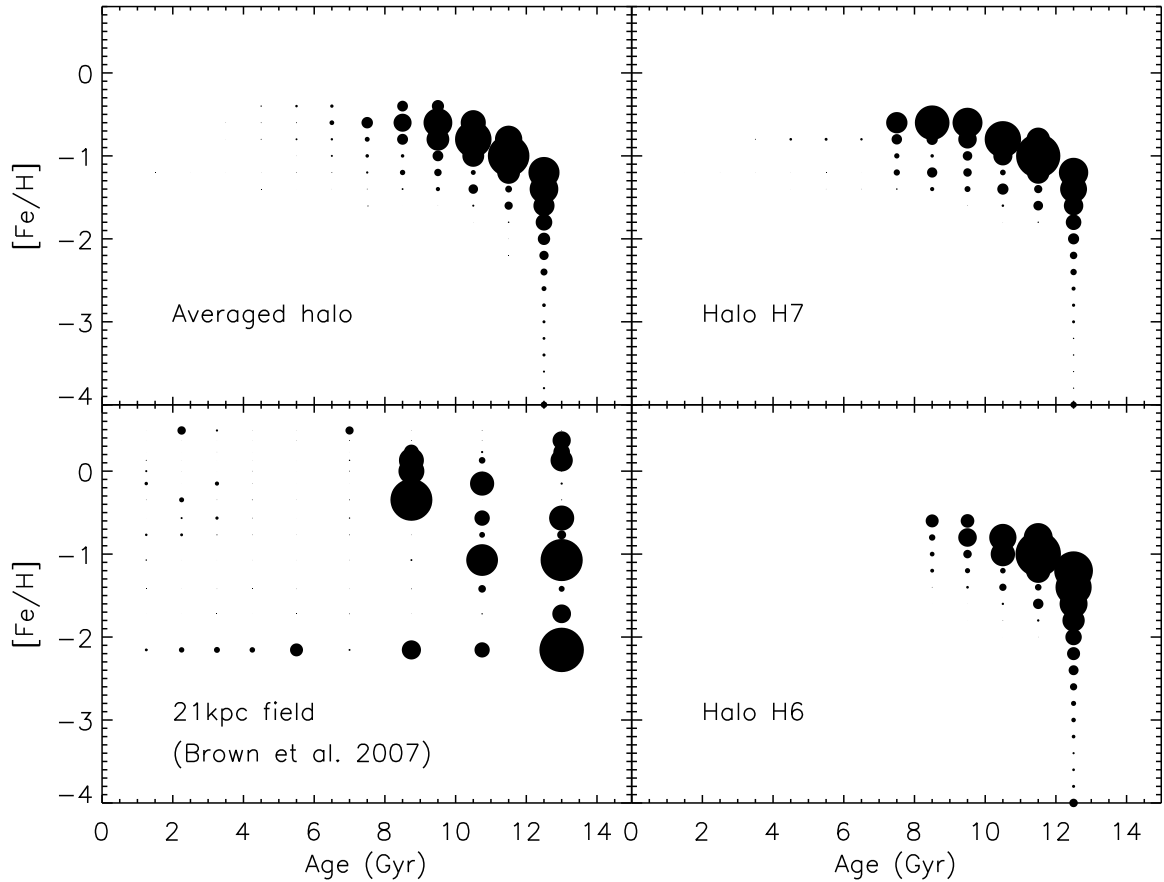}
\caption{\label{fig:inner_comp}{Comparison between the halo field of
	 \citet{brown07} and simulated inner halo fields. Top left panel
	 shows the averaged halo; top and bottom right panels show two
	 individual halos with a protracted and short and mass accretion
	 histories, respectively. Bottom left panel shows the observational
	 data (Courtesy of Tom Brown). The symbol size in the simulated fields
	 are as indicated in Figures \ref{fig:agefe_master} and 
    \ref{fig:agefe_inner_outer}. For the observational data, the area
	 of the circles is proportional to the number of stars in each
	 isochrone group (see \citet{brown07}).}}
\end{figure*}

\subsection{The M31 Halo}
\label{sec:comp_obs_halo}

\citet{brown06a,brown06b,brown07} have derived the star formation histories at various locations in the outskirts of M31. The resulting ages and metallicities in these fields can be compared directly to our model predictions. A concern is that some of Brown's fields lie along the minor axis of M31, a region that has been shown to contain stars from multiple sources, including the disk and GSS \citep{ibata07}. The most recent field analyzed lies at 21~kpc along the minor axis and is expected to have an insignificant contribution from thin disk stars or from the giant extended disk structure detected by \citet{ibata05}  (for evidence against contamination from the latter component, see \citet{gilbert07}; also, note that the extended disk component may have an accretion origin as well \citep{penarrubia06}). On the other hand, the 21~kpc halo field may still be contaminated by stream debris \citep{gilbert07}.  

We note, however, that from the point of view of hierarchical models one cannot easily differentiate between a ``true'' or ``clean'' halo and a halo contaminated by debris. In this context, stellar halos are expected to form through the superposition of tidal debris and show various degrees of phase-mixing. Therefore the only concern in our comparison is not to mistake a field that is dominated by a given stream for one that is representative of the whole halo. The cautionary remark regarding our further comparison with the $21$~kpc field of \citet{brown07} is that the presence of a cold stream in the inner halo field appears indeed to be atypical in our simulations. Whether this observed field is representative of the whole inner M31 halo, or of a typical Milky Way-type inner halo, remains to be decided by future observations.

Our strategy for further comparisons with the observations is to present averaged results rather than selected, well-matched fields. In this respect, our results do not emphasize the dispersion in stellar population data that can exist among individual halo fields. However, these differences are not expected to be as large in the inner halo as in the less phase-mixed, outer one \citep{font06b}.

\subsubsection{The $21$~kpc Halo Field}
\label{sec:comp_obs_21kpchalo}

Figure \ref{fig:inner_comp} shows a comparison between the $21$~kpc halo field of \citet{brown07} and our averaged inner halo region (bottom and top left panels, respectively). In the right panels we also show the AMR of two individual halos from our sample that have mass accretion histories that bracket the averaged one (halo H7 has a protracted mass accretion history, whereas halo H6 has a short one. Halo H7 is also $\sim 0.5$~dex more metal-rich than halo H6).

The dominant presence of old stars in both observations and simulations is not surprising, as inner halos are expected to form within the first few Gyr from the assembly history \citep{bullock05}. The existence of intermediate-age stars in the inner halos is a new finding that can provide important constraints on the models of halo formation. \citet{brown06b} estimate that $\sim 40\%$ of the stars in the field $11$~kpc along the minor axis in M31 are less than $10$~Gyr old, while the mean age in that field is $9.7$~Gyr. \citet{brown07} find that the field at $21$~kpc along the minor axis has significantly less stars with ages less than $10$~Gyr (the fraction is about half of that in the inner field), and virtually no stars with ages less than $8$~Gyr. The distribution of populations in the latter field seems to be more into agreement with our models, which predict that about $20\%$ of populations in the inner halo have ages less than $10$~Gyr.

The AMRs of the simulated halos show non-negligible populations with
intermediate metallicities $-2 <$\FeH $ <-1$ and ages younger than $10-11$~Gyr. Although these populations represent only a minor contribution ($<1\%$, by mass) to the entire stellar halo, they can provide important information about the halo formation history. In our models, stars with these ages and metallicities predominantly come from lower-mass, metal-poor halo progenitors, and hence are a probe of the accretion of dwarf galaxies at the lower end of the mass function. (Note: the lower metallicity of low mass halos follows because we imposed that dwarfs accreted today follow the $M_{*} -$ Z relation \citep{dekel86,dekel03}.) 

The old and metal-poor stars with ages $>10$~Gyr and \FeH $<-2$,
located in the lower right corners of the AMRs, are good tracers of the
early stages of star formation in galaxy halos or in their
progenitors. Both our models and the observations show non-negligible
fractions of such metal-poor stars (note: the \citet{brown07} AMR shows
only stars with \FeH $>-2$, but this may only be an artifact of the
limited choice of metallicities in the isochrone dataset). At the
moment there is a great interest in finding the most metal-poor stars
in halos of galaxies and in understanding their origin. Apart from the 
evidence that the M31 halo has some very metal-poor stars 
\citep{brown04,brown06b,brown07}, the searches also target the Milky Way halo \citep{beers05}. In our models most of the stellar populations form in small sub-halos at high redshift and enter the main halo in relatively massive satellites. Improved models that track the early episodes of star formation and metal enrichment in more detail will be able to provide better constraints on the origin of this type of populations (see Tumlinson et al, in prep.).

Interestingly, the observed AMR of \citet{brown07} contains some stars that are
more metal-rich than the stars of the same age in our simulated
halos (see, in particular, the [8~Gyr,12~Gyr] age interval). This rapid
enrichment is indicative of stars originating in a fairly massive
satellite, of $\sim 10^{9}$ or a few $10^{8} \, M_{\odot}$ in stellar
mass (see, for example the AMRs of massive satellites in Figure
\ref{fig:agefe_streams}). The reason that our simulated halos do not
show such signatures is that most of the progenitors of these well-mixed
regions have stellar masses of $10^{7-8} \, M_{\odot}$ or
less. According to our models, the accretion of a $M_{*} \sim 10^{8-9} \,
M_{\odot}$ satellite is a rare event it it were to occur more than
$8$~Gyr ago. A more recent accretion of such a massive satellite is
possible, however the chances that the satellite is fully disrupted
such that it mixes with the rest of the halo are small, unless its
orbit is extremely radial. A possible explanation for the origin of
these old and metal-rich populations is that they originate in the
GSS (recall that the 21~kpc field of \citet{brown07} {\it is}
contaminated by the GSS; and that the progenitor of the GSS is believed
to have been fairly massive and on a radial orbit. See also
the discussion of the GSS data in sections \S \ref{sec:comp_obs_streamAMR} and \S \ref{sec:sfr_orbit}). 

There are also a number of discrepancies between our models and the
observations of \citet{brown07}. The stars with ages $< 5$~Gyr and \FeH
$>-1$ in the top left corner of the observed AMR cannot be matched by our
models. It is, however, unlikely that these extremely young and
metal-rich stars belong to the halo. We hypothesize that they are
either caused by a very minor contamination by the disk in this field,
or that they belong to a recently accreted stream. The reason that our 
simulated AMRs do not have this young and metal-rich component is
because, {\it on average}, streams of satellites accreted less than
$6$~Gyr ago are not found in this inner regions of our simulated halos.

Other more extreme types of populations inferred by
\cite{brown06b,brown07} in the halo of M31 are also not easily retrieved in
our models $-$ for example, the trace populations of young and
extremely metal-poor stars (ages $<10$~Gyr and \FeH $<-2$) in the
bottom left corner of the observed AMR or the old and metal rich stars 
(ages $>10$~Gyr and \FeH $>-0.5$) in the top right corner of the same
diagram. The reason for this mismatch may be that the modelling of
the gas and stellar physics in our models is too simplistic to account
for these populations. However it is difficult to understand what type
of satellite systems would have such unusual paths of chemical
evolution. In particular, if the young stars with metallicities \FeH
$<-2$ do indeed have a merger origin, it implies that the progenitor
satellites either formed only recently and in relative pristine
environments, or that the mixing of metals inside these satellites was
very inefficient.  We note, however, that while the presence of these
extreme populations was derived from the \cite{brown06b} analysis, the
authors note themselves that they may not actually be required.
Indeed, \cite{brown06b} showed that old metal rich stars could easily
be excluded from their model without diminishing the goodness of
fit. Furthermore, they point out that what appears as a young metal
poor population in their CMD could at least partially be explained as
a population of blue stragglers.

\begin{figure*}[t]
\includegraphics[width=16cm]{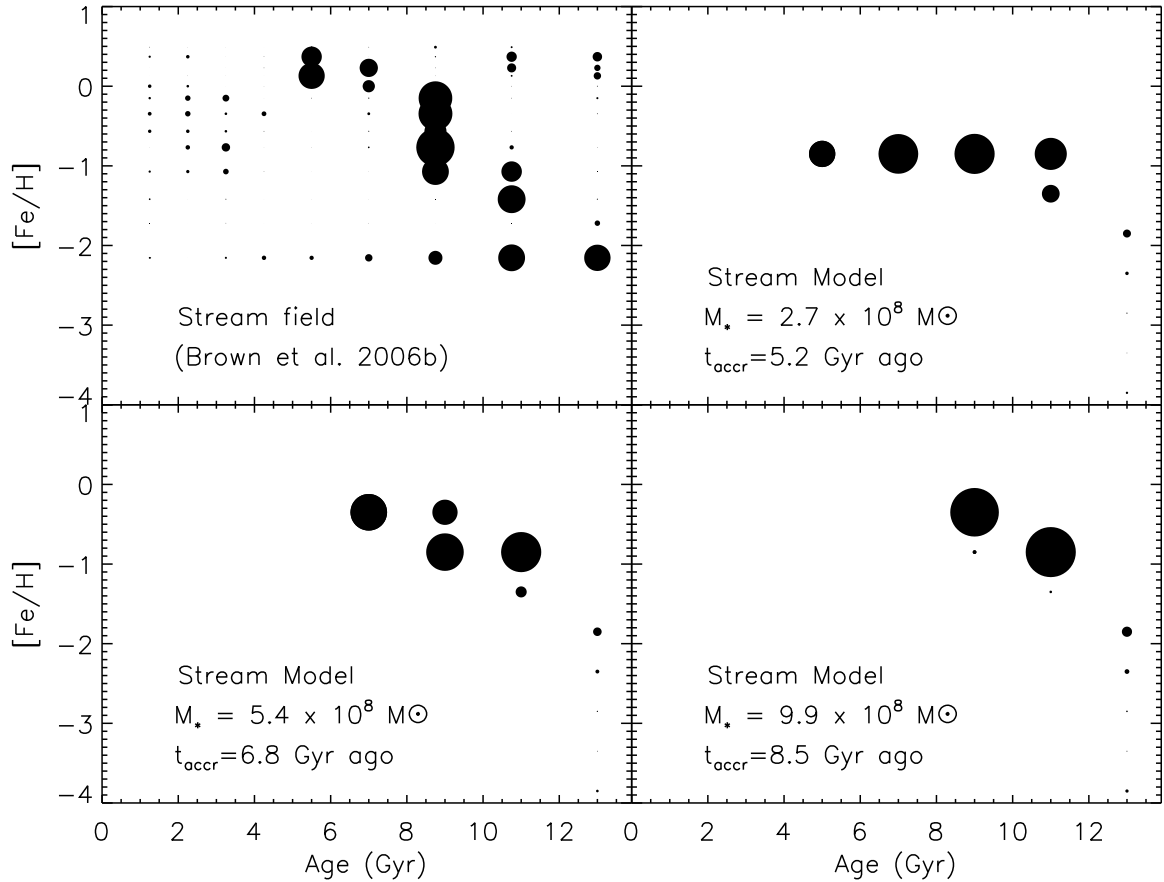}
\caption{\label{fig:stream_comp}{Comparison between the GSS field of
	 \citet{brown06b} and several simulated streams. (Observational data
	 are courtesy of Tom Brown). The symbol size in the simulated streams
	 are as indicated in Figures \ref{fig:agefe_streams}. For the
	 observational data, the area of the circles is proportional to the
	 number of stars in each isochrone group (see \citet{brown06b}).
}}
\end{figure*}

\subsubsection {The Outer Halo}
\label{sec:comp_obs_outerhalo}

Figure \ref{fig:agefe_inner_outer} also shows that the metal-poor populations (\FeH $<-2$) extend out to large distances in the halo, in agreement with recent observations in M31 which detect an underlying metal-poor halo component out to $\sim 150$~kpc \citep{chapman06,gilbert06,kalirai06,ibata07}. 

We find that the ratio of metal-poor stars to metal-rich stars is higher in the outer halo than in the inner one\footnote{This result does not imply that the {\it absolute} number of metal-poor stars in outer halo should be higher than in the inner halo. The density of metal-poor stars is expected to decrease with distance from the center of the galaxy (see \citet{font06b}).}. This result appears to be in agreement with observations that find that the overall metallicity of stars decreases with radius in the M31's spheroid \citep{bellazzini03,kalirai06,brown07}. In contrast, the density profiles derived by \citet{irwin05,ibata07} for the metal-poor and metal-rich populations in M31 appear very similar, suggesting that the ratio of these populations is fairly constant from inner to the outer halo. 

We also predict that the outer regions of the halos should have, on
average, younger ages than the inner ones, as a consequence of the
inside-out growth of these halos. The two fields examined by
\citet{brown07} at $11$~kpc and $21$~kpc along the minor axis of M31
show an inverse trend. However, it is likely that the excess of younger
stars detected in their inner field is the result of contamination by
the disk, rather than a genuine trend (however, note that Brown and
collaborators argue against disk contamination in the 11~kpc field). It is certainly desirable to have a detailed analysis of {\it both} ages and metallicities in more fields across the M31 halo before drawing a definitive conclusion about the age and metallicity gradients in this system. 

\subsection{The Giant Southern Stream}
\label{sec:comp_obs_stream}

\subsubsection{Stellar Populations}
\label{sec:comp_obs_streamAMR}

Figure \ref{fig:stream_comp} shows the AMR of the GSS field derived by \citet{brown06b} alongside the AMRs of several simulated streams. We have chosen for this comparison only streams that have massive progenitors, $M_{*}>10^{8} M_{\odot}$, as several observational and theoretical studies indicate that this is the case of the GSS progenitor \citep{ibata04,brown06a,font06c}. For example, we find that only progenitors with total stellar masses greater than $\sim 5 \cdot 10^{8} M_{\odot}$ are able to match the most metal-rich portion of the observed AMR.  Among these massive satellites, we find that those accreted $\sim 6-7$~Gyr ago provide a better fit to the observed AMR (eg. the bottom left panel in Figure \ref{fig:stream_comp}) and also result in tidal debris of comparable surface brightness as the GSS, $\mu_{V} \sim 30$ \magarc \, \citep{ibata01}. (A more detailed discussion about constraining the time of accretion will be given in \S \ref{sec:sfr_orbit}). 
 
The AMRs of satellites accreted $6-7$~Gyr ago also have about $70\%$ of their stars younger than $10$~Gyr, in agreement with what has been found in the GSS \citep{brown06b}.

Not all stellar population data can be matched by our models. The metal-poor populations (\FeH $<-2$) are not as significant in the simulated AMR than in the stream field of \citet{brown06b}. Also, as in the case of main halos, the most extreme populations detected in the GSS field, such as the top right, top left and bottom left corners of the \agefe grid of \citet{brown06b} are inconsistent with our chemical evolution models.

As discussed before, the mismatch between the observations and the
models can be attributed, in part, to the limitations of our current
modelling of individual streams. Additionally, the observational data
in the GSS field may contain stars from other components, such as the
disk, other tidal debris, or the background halo. For instance, 
\citet{brown06b} estimate that about $25\%$ of the stars in the stream
field could belong to the underlying halo.  In addition, as previously 
discussed in \S \ref{sec:comp_obs_21kpchalo}, some of these extreme 
populations, such as the old and metal-rich ones, are not necessarily 
required for a good fit to the observations (although the young and
metal-poor ones seem to be required).

\subsubsection{Constraining the Time of Accretion}
\label{sec:sfr_orbit}

Although the constraints on the mass of the GSS progenitor are fairly good, the time of accretion is largely uncertain\footnote{The time of accretion is defined here as the time when the satellite enters the virial radius of the host galaxy.}. Here we discuss the possibility of putting some constraints on the latter by using our star formation history models.

Even though stellar population data contain direct information only
about the star formation history of the progenitor galaxy, indirectly,
they can also provide some constraints on the external mechanisms that
can shut off the star formation. The accretion of the satellite onto
the main halo is such a mechanism that may affect the star formation
process within the satellite, slowing it down or stopping it
completely. \citet{brown06b} suggested that the cessation of star
formation in the GSS $\sim 6-7$~Gyr ago might have coincided with the accretion of the progenitor satellite onto the M31. For example, if the satellite were accreted earlier than $7$~Gyr ago, ram-pressure stripping along this highly eccentric orbit could have removed the gas reservoir and therefore stopped the star formation earlier than it is observed. Similarly, if the satellite were accreted more recently, star formation would have continued several Gyrs after what is observed, unless some intrinsic mechanism would have stopped it before the accretion. 

However, a time of accretion of $6-7$~Gyr ago is at odds with current numerical modelling of the satellite orbit that fits the kinematic data. The simulations find that the progenitor satellite has spent less than $1$~Gyr orbiting from the farthest location along the stream until the current pericenter \citep{ibata04,fardal06,font06c,geehan06,fardal07}, and assuming that farthest edge of the stream is the point where the satellite was accreted, it implies that the satellite must have been accreted within the last $\sim 1$~Gyr.

Although the stellar population data cannot provide stringent constraints on the dynamical evolution of the satellite, the arguments above point towards a possible conflict with the numerical estimates. We examine below two possibilities: first, that the progenitor was accreted recently ($<1-2$~Gyr ago), but the star formation stopped several Gyr prior to that; and second, that the satellite was accreted $6-7$~Gyr. 

\begin{figure}
\centering
\includegraphics[width=9cm]{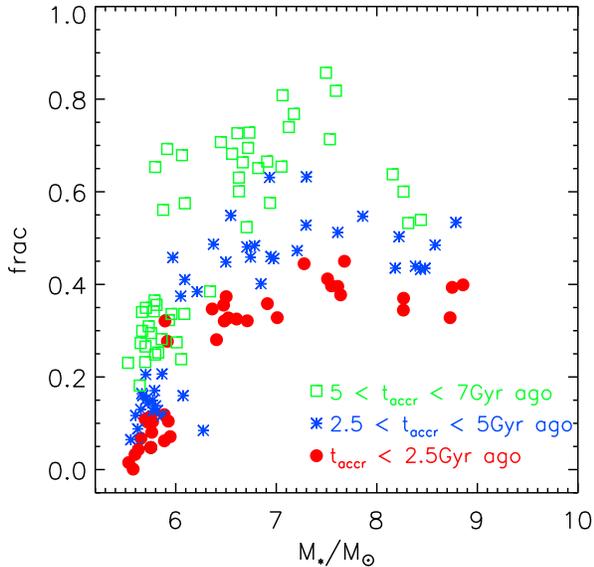}
\caption{\label{fig:sf_stop}{The fraction of stars formed in the $4$~Gyr before the time of accretion in simulated satellites, function of the total stellar mass of the satellite. Different symbols represent different accretion times: less than $2.5$~Gyr ago (red circles), between $2.5$ and $5$~Gyr (blue star symbols), and between $5$ and $7$~Gyr ago (green squares).}}
\end{figure}

In the first case, the question is how likely is that star formation
stops several Gyrs prior to the accretion, particularly in a fairly
massive satellite as it is inferred to be the case for the progenitor
of the GSS. In Figure \ref{fig:sf_stop} we investigate this scenario
quantitatively, by selecting all satellites accreted less than $7$~Gyr
ago and plotting the fraction of stars formed in the $4$~Gyr time
interval before accretion. The satellites are divided in three
categories depending on their time of accretion: satellites accreted
about $5-7$~Gyr ago, satellites accreted recently ($<2.5$~Gyr), and
satellites accreted at intermediate times. The fraction of stars is
expressed function of the total stellar mass of the satellite at the
time of accretion. For more massive satellites, $M_{*} > 10^{7} \,
M_{\odot},$ $-$ as it is assumed for the progenitor of the GSS $-$ at
least $40\%$ of all their stars are formed within the $4$~Gyr before
accretion. The implication is that, if the progenitor were indeed
accreted within the last $1-2$~Gyr, the observations should detect a
$\sim 40\%$ fraction of young stars (ages $<6$~Gyr). Current
observations, however, do not detect such a high fraction of young stars \citep{brown06b}. 

Therefore, the results based on our models seem to concur that, for a
massive satellite, a time of accretion of $1-2$~Gyr ago is difficult to
reconcile with a cessation of star formation $4$~Gyr prior to that. The
caveat, of course, is that the star formation prescription in our
models may be too simplistic. The alternative is that the progenitor of
the GSS was indeed accreted $6-7$~Gyr ago. In this case, it is likely
that the satellite orbited more than once in the M31 halo, as the time
taken to complete an orbit with the current kinematic properties is
much less than $6-7$~Gyr (see the numerical simulations of
\citet{ibata04,fardal06,font06c,geehan06,fardal07}). Also, it is also
unlikely that a tidal stream that detached $7$~Gyr ago would still
survive today as a coherent feature. Therefore, if the satellite was
accreted $6-7$~Gyr ago, the stars in the current stream must have been
tidally stripped from the parent halo relatively recently. It is possible that a resolution between the results of numerical modelling and stellar population data can be achieved if the starting point of simulations is moved beyond the edge of the currently detected stream. Extrapolating the numerical parameter space in regions which are not constrained by the data would be difficult, but certainly worth doing. The task of constraining the orbital parameters could be greatly eased if future observations find a continuation of the stream beyond the current edge, either along this orbit or on a  previous one.

Our current simulations can provide some predictions on how faint the
streams of satellites that undergo multiple orbits should be. The small
panels at the bottom of the Figure \ref{fig:short} illustrate the
surface brightness of a few streams formed when the parent satellite is
at its first and then second passage inside the main halo. The streams
are brightest when they just form, then faint with time. Therefore
the streams that formed in a recent passage of a satellite inside the main
halo are expected to be brighter than the streams that survive from
previous a orbit of that satellite (however, the effect is attenuated
by the fact that more stellar material is stripped at earlier passages
of the satellite, therefore the {\it initial} surface brightness of the
streams is higher for the streams formed earlier). We find that while
the brighter (i.e. more recently detached) streams have an average surface 
brightness of $\sim 25-26$ \magarc, the fainter counterparts have, 
on average, $\mu_{V} \sim 26-27$ \magarc. These surface brightness
levels should be achievable with current detection limits.

\section{Conclusions}
\label{sec:concl}

In this paper we used a set of simulations of stellar halos of $L_{*}$ galaxies to study the composition of their stellar populations and provide comparisons with the observations. The test case for our comparisons was M31, given the wealth of data that exists for this galaxy. However, we emphasize that the theoretical results derived here are applicable to any large $\sim L_{*}$ galaxy and therefore provide a framework for interpreting the substructure of such galaxies that will be uncovered in upcoming surveys.

The main conclusions of our study are as following:

- The global stellar populations of halos assembled hierarchically will vary according to their different accretion histories. For example, the presence of metal-rich debris in a galaxy like M31 suggests a history that included the accretion of at least one massive satellite. 
Conversely, the more metal poor halo of the Milky Way is evidence that
its progenitor satellites were less massive than those of M31.

-  The homogeneous high metallicity in the inner M31 halo found by
   \citet{brown07} can be naturally explained in the context of
   hierarchical merging. According to our models the homogeneity of
   stellar populations occurs because the inner halo is dominated by a
   few massive satellites, $M_{*} \simeq 10^{7-8} M_{\odot},$ which had
   very similar star formation and chemical evolution histories. The
   predicted imprint of these satellites in the inner stellar halo is
   the presence of old stars with \FeH $\simeq -1 \div 0$.

- The outer halos in our models tend to have a larger spread in the ages and metallicities of stellar populations than the inner parts. This diversity may not be apparent in current observations because the spread is contributed in part by a multitude of low-surface brightness features. Ongoing deep observations in the M31 should be able to test this prediction.

- We have investigated the origin of high surface debris in different regions of the halo. The debris in the outer $R>10$~kpc halo typically originates in single massive satellites. This result holds regardless of whether the debris is kinematically cold or if it is diffuse in phase-space. The outer regions give the best opportunity of reconstructing the formation history of stellar halos, as their stellar populations retain information about the nature of the mergers in a non-degenerate form.

- The origin of bright debris in the inner $10$~kpc halo is more varied. Here tidal debris from a single massive progenitor or superposition of debris from several large or small satellites can result in similar surface brightness levels. Disentangling the complex formation histories of these fields is possible, provided that the interpretation of observational data takes into account the predictions of hierarchical models.

- The brightest tidal debris (i.e. that can be observed with current
  surface brightness limits) still orbiting around M31 or Milky
  Way-type galaxies should be metal-rich. Therefore it should not be
  expected to resemble the characteristically old and metal-poor
  stellar populations of dwarf spheroidal satellites, nor those of the
  smooth underlying halo.

- We illustrated the case of GSS in M31, whose progenitor was a massive dwarf galaxy accreted more recently than the main progenitors of the halo. The presence of intermediate-age stars and the absence of young stars suggest that the satellite could have been accreted as long ago as $6-7$~Gyr. If the satellite was accreted more than $1-2$~Gyr ago, a continuation of the stream should exist in the halo of M31, about $1-2$ \magarc \, fainter than the currently observed one.

\acknowledgements
We thank the referee for a thorough reading of the manuscript and for
useful suggestions that improved the presentation of our paper. We wish 
to thank Tom Brown for providing us with observational data for
the halo and stream fields in M31. We also thank
Tom Brown, Henry Ferguson and Karrie Gilbert for helpful conversations
and the Aspen Center for Physics for hospitality during the writing of
this paper. Support for programs GO9458 and GO-10128 was provided by
NASA through a grant from the Space Telescope Science Institute, which
is operated by Association of Universities for Research in Astronomy,
Inc., under NASA contract NAS 5-26555. ASF and KVJ's contributions were
supported through the NSF CAREER award AST-0133617. ASF is supported by
a PPARC Fellowship at the Institute for Computational Cosmology in
Durham.  AMNF is supported by a Marie Curie Excellence Grant from the
European Commission under contract MCEXT-CT-2005-025869. BER gratefully
acknowledges the support of a Spitzer Fellowship through a NASA grant 
administered by the Spitzer Science Center.  J. T. gratefully
acknowledges the generous support of Gilbert and Jaylee Mead for their 
namesake fellowship in the Yale Center for Astronomy and
Astrophysics. PG acknowledges support from NSF grants AST-0307966 and  
AST-0507483 and NASA/STScI grants GO-10265.02 and GO-10134.02.

\bibliographystyle{apj}

\end{document}